\documentclass[12pt,epsfig,epsf]{article}
\usepackage{graphicx}
\def\bcc{\begin{center}}
\def\ecc{\end{center}}

\def\dm2{\rm{\Delta m^2}}

\def\s2tw{\rm{ sin ^2 \theta _W }}
\def\am241{\rm{ ^{241} Am }}
\def\u238{\rm{ ^{238} U }}
\def\th232{\rm{ ^{232} Th }}
\def\k40{\rm{ ^{40} K }}
\def\th232{\rm{ ^{232} Th }}
\def\u238{\rm{ ^{238} U }}
\def\cs137{\rm{^{137} Cs }}
\def\ba133{\rm{^{133} Ba }}

\def\s2tw{\rm{ sin ^2 \theta _W }}

\def\ke10{\rm{\kappa_e}}

\begin{document}

\bcc
{\Large Estimation of Impact Parameter on event-by-event basis in Nuclear Emulsion Detector}
\ecc

\bcc
V. SINGH\footnote{vsingh@phys.sinica.edu.tw, Present Address: Institute of Physics, Academia Sinica, Taipei, Taiwan (ROC).}\\
Department of Physics, Banaras Hindu University, Varanasi, India\\
B. BHATTACHARJEE, S. SENGUPTA\\
Department of Physics, Gauhati University, Guwahati, India\\
A. MUKHOPADHYAY\\
Department of Physics, North Bengal University, New Jalpaiguri, India\\
\ecc



\bcc
Abstract
\ecc
Photographic Nuclear Emulsion Detector (PNED) has been in use in nuclear and 
particle physics experiments from the begining, often as the major detector system. 
However, direct measurement of impact parameter in this detector does not seem  
possible due to some limitations. This paper discribe a simple yet strong method 
to estimate the impact parameter of events on event-by-event basis. Though this 
method is develpoed specifically for the photographic nuclear emulsion detector, 
we envision that it should also be applicable to other multi-target detector systems.
\vskip 0.2cm
{\bf Keywords:} NUCLEAR REACTION Impact parameter, photoemulsion method, relativistic 
nuclear collisions.

\section{Introduction}

A study of relativistic nucleus - nucleus collision is an important 
tool to produce and investigate highly dense nuclear matter in the 
laboratory {\bf [1-3]}. Based on straightforward geometrical 
considerations, theoretical models predict that the size of the dense 
nuclear matter zone produced in collisions depends strongly on the 
impact parameter (b) i.e., the transverse distance between the center 
of mass of the projectile and the target nucleus {\bf [4]}. It is thus 
very important to sort out the collisions according to their centrality. 
The impact parameter, which characterizes the initial state, is not a 
directly measurable quantity. Thus, it is necessary to find out an 
observable that strongly correlated with it. The simplest observable 
one can think of is the total charged particle multiplicity of an event 
in case of nuclear emulsion detector. We can also try to use the shower 
(mostlly pions) particle or projectile's proton multiplicity as an 
observable but it does not show a strong correlation with impact 
parameter due to different size of the targets. In emulsion detector; 
emulsion provides medium to the projectile as well as targets. Emulsion 
mainly composed of H, CNO and Ag(Br). The variation in target size could 
create big confusion in identifcation of events having b = 0 and events 
having impact parameter $b>0$ of the bigger size projectile with H and Ag(Br).
In this paper, we have developed a simple method for estimation of impact 
parameter on an event-by-event basis, which allows one to translate the 
qualitative estimate of impact parameter from total multiplicity into a 
qualitative one. A quantitative estimate is very convenient in order to 
present consistent results obtained in various experiments. We will also 
discuss some basic characteristics of the interactions with respect to 
the impact parameter. Some of the charateristics are studied in different 
energy intervals like high, Mid., and low energy.

\section{Experimental Details}

Nuclear emulsion is a detector composed of silver halide crystals immersed in a gelatin matrix {\bf [5-7]} 
consisting mostly of hydrogen, carbon, nitrogen, oxygen, silver and bromine while a small percentage 
of sulfur and iodine are also present as shown in {\bf table 2}. In the present experiment, we have employed a stack of high 
sensitive {\bf NIKFI BR-2} nuclear emulsion pellicles of dimensions 9.8$\times$9.8$\times$0.06~$cm^{3}$, exposed horizontally 
to $^{84}Kr$ ion at a kinetic energy of around 1 GeV per nucleon. The exposure has been performed at 
Gesellschaft fur Schwerionenforschung {\bf (GSI)} Darmstadt, Germany. The events have been examined and 
analyzed with the help of a {\bf LEITZ (ERGOLUX)} optical microscope having total magnification of 2250X 
and measuring accuracy of 1 $\mu$m. In order to obtain an unbiased sample of events, an along-the-track 
scanning technique has been employed {\bf [8]}.

\begin{table}
\begin{center}
\caption{Mean Free Path of different projectiles in nuclear emulsion.}
\begin{tabular}{cccc}
\hline
{\bf Projectile}&$\bf Energy~(A~GeV)$&$\bf Mean~Free~Path~(cm.)$&{\bf Ref.}\\
\hline
$^{4}He$&2.1&21.80$\pm$0.70&{\bf [9]}\\
$^{12}C$&2.1&13.80$\pm$0.50&{\bf [9]}\\
$^{14}N$&2.1&13.10$\pm$0.50&{\bf [9]}\\
$^{16}O$&2.0&12.60$\pm$0.50&{\bf [10]}\\
$^{56}Fe$&1.7&7.97$\pm$0.19&{\bf [11]}\\
$^{84}Kr$&1.0&6.76$\pm$0.21&{\bf [Present work]}\\
$^{139}La$&1.2&5.18$\pm$0.30&{\bf [12]}\\
$^{197}Au$&1.0&5.60$\pm$0.26&{\bf [13]}\\
$^{238}U$&1.0&3.67$\pm$0.12&{\bf [14]}\\
\hline
\end{tabular}
\end{center}
\end{table}

\begin{table}
\begin{center}
\caption{The chemical composition of NIKFI BR-2 emulsion {\bf[15]}.}
\begin{tabular}{ccccccc}
\hline
{\bf Element}&$\bf ^{1}H$&$\bf ^{12}C$&$\bf ^{14}N$&$\bf ^{16}O$&$\bf ^{80}Br$& $\bf ^{108}Ag$\\
\hline
No of atoms/cc~$\times$~$10^{22}$&3.150&1.410&0.395&0.956&1.028&1.028\\
\hline
\end{tabular}
\end{center}
\end{table}

The interaction mean free path $(\lambda)$ of $^{84}Kr$ in nuclear emulsion has been determined and found to be 
6.76$\pm$0.21 cm. Our collaborators {\bf (DGKLMTV Collaboration) [16]} found a value of mean free path 
($\lambda$) 7.10$\pm$0.14 cm. consistent, within the experimental error, with our value. 
We have tabulated the mean free path of different emulsion experiments at similar beam energy in {\bf table 1}.
From this table, we may conclude that the mean free path decreases with increasing beam mass number at similar energy.
The mean free path value obtained in our experiment is well fitted in this trend i.e., during event scanning, 
we picked up all genuine events according to our event selection criteria and our criteria of event selection 
is also right. For the present work, we used 1197 events scanned by line scanning method and additional 162 events 
were picked up by volume scanning mehtod. The grain density of a singly charged particle passing in the same emulsion 
at extreme relativistic velocity is called the minimum grain density ($g_{min}$). In this experiment, 
its measured value is equal to 28$\pm$1 grains per 100~$\mu$m. The $^{84}Kr$ beam stops within a pellicle.
Since, the beam energy decreases as it goes from the entrance edge, we have divided each plate in three 
major energy intervals where the beam has energy in the range 0.95 - 0.80 {\bf(High Energy)}, 0.80 - 0.50 
{\bf(Middle Energy)}, and below 0.50 {\bf(Low Energy)} A GeV, respectively.

The mean number of fully developed and well separated grains per unit length is called the grain density 
{\bf g}. It is a measure of the rate of ionization loss. The grain density of a track corresponds to a 
particular specific ionization but its actual value depends on the degree of development of the emulsion 
and the type of the emulsion used. It is therefore, necessary to introduce another quantity called 
normalized grain density which is defined as $g^{*} = g / g_{min}$. Here $g$ is the observed grain density. 
All charged Secondaries emitted or produced in an interactions are classified in accordance with their 
ionization, range and velocity into the following categories:

{\bf (a) Shower tracks ($N_{s}$):} These are freshly created newly produced charged particles with  
$g^{*}<1.4$. These particles have relative velocity $\beta > 0.7$. For the case of a proton it means 
energy of $E_{p} >400$ MeV. They are mostly fast pions with a small admixture of kaons and of 
released protons from the projectile which have undergone an interaction. These conditions ensure that 
showers are filtered from the fragments and knockout protons of the target.

{\bf (b) Grey tracks ($N_{g}$):} Particles having ionization in the interval 1.4 $< g^{*}< 6.0$ and 
range $>3$ mm are defined as greys. These particles having relative velocity $0.3 < \beta <0.7$. 
They are generally knocked out protons of targets having energy $30 < E_{p} <400 $ MeV but also 
admixture of deutrons, tritons and some slow mesons.

{\bf (c) Black tracks ($N_{b}$):} Particles having range $< 3$ mm from interaction vertex from which 
they originated and $g^{*} >6.0$. This corresponds to a relative velocity $\beta <0.3$ and a proton 
with energy $E_{p} <30$ MeV. Most of these are produced owing to evaporation of residual target nucleus.

The {\bf heavily ionizing charged particles} ($N_{h}$ = $N_{g}$ + $N_{b}$ ) are parts of the target 
nucleus and are also called target fragments. 

{\bf (d) Projectile Fragments ($N_{f}$):} These are the spectator parts of the projectile nucleus 
with charge $Z\ge 1$ having velocity close to the beam velocity. The ionization of projectile fragments 
(PFs) is nearly constant over a few mm and emitted within a highly collimated forward narrow cone whose 
size depends upon the available beam energy.

The forward angle is the angle whose tangent is the ration between the average transverse momentum of 
the projectile fragments to the longitudinal momentum $(p_{L})$ of the beam. Taking $p_{L}$ as the 
beam momentum itself, i.e., $\theta_{F} = tan^{-1} (p_{t} / p_{L}) = \sim 9^{o}$ in this experiment.
The PFs are further classified into three categories as follows:

{\bf (i) Heavy Projectile Fragments ($N_{f}$):} PF's with charge $Z\ge 3$.

{\bf (ii) Alpha Projectile Fragments ($N_{\alpha}$):} PF's having charge Z=2.

{\bf (iii) Singly charged relativistic Projectile Fragments ($N_{f_{Z=1}}$)}.

Since these PF's have velocities nearly equal to the initial beam velocity, their specific ionization 
may be used directly to estimate their charge.

The {\bf total multiplicity of the secondary charged particle ($N_{ch} or M$)} is taken as the sum of 
all charged particles emitted or produced in an interactions 
$(N_{ch} = N_{s} + N_{h} + N_{f} + N_{\alpha} + N_{f_{Z=1}})$. 

\begin{figure}
\center
{\bf a)}
\includegraphics[height=6.6cm,angle=0]{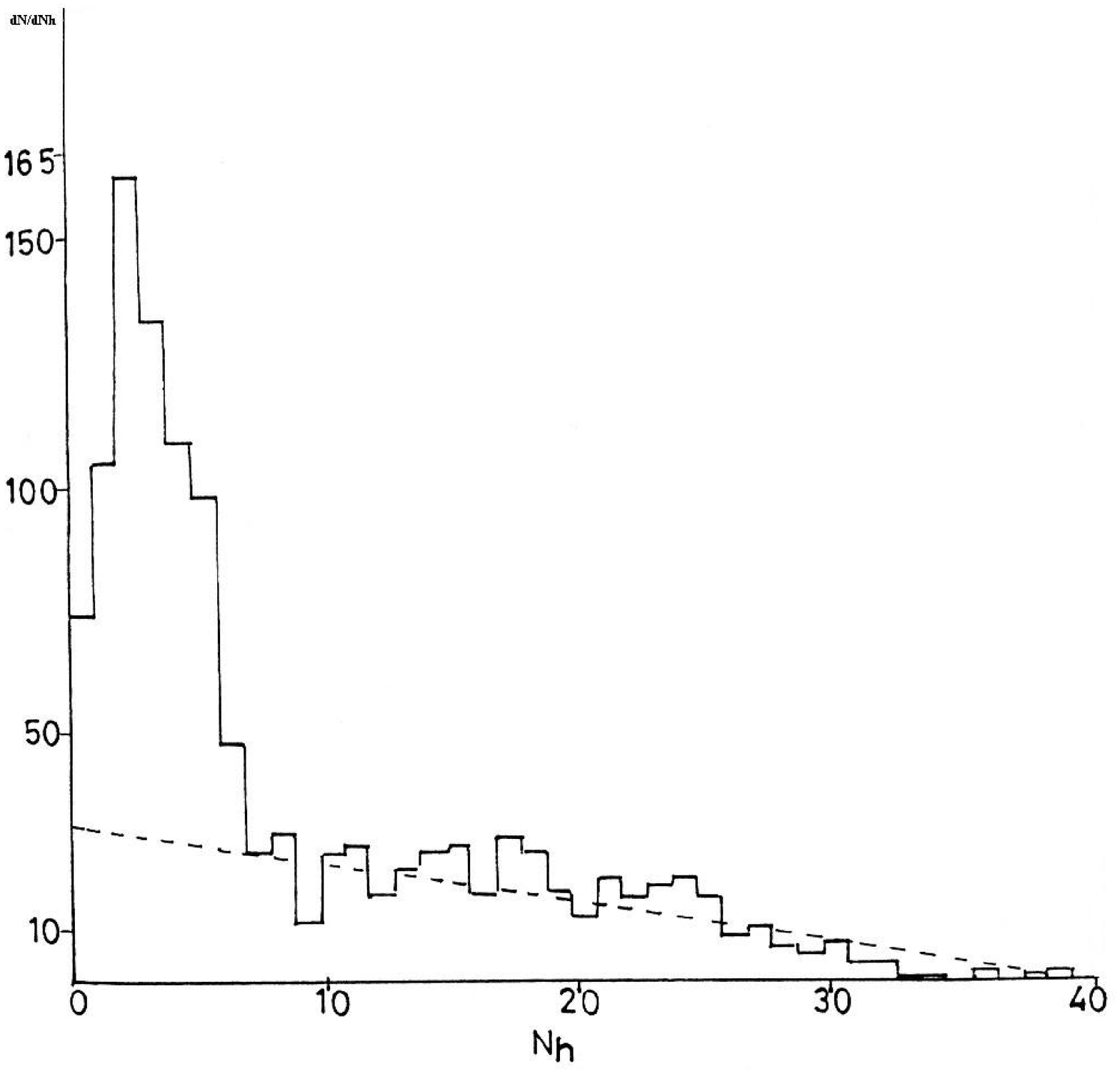}
{\bf b)}
\includegraphics[height=7cm,angle=0]{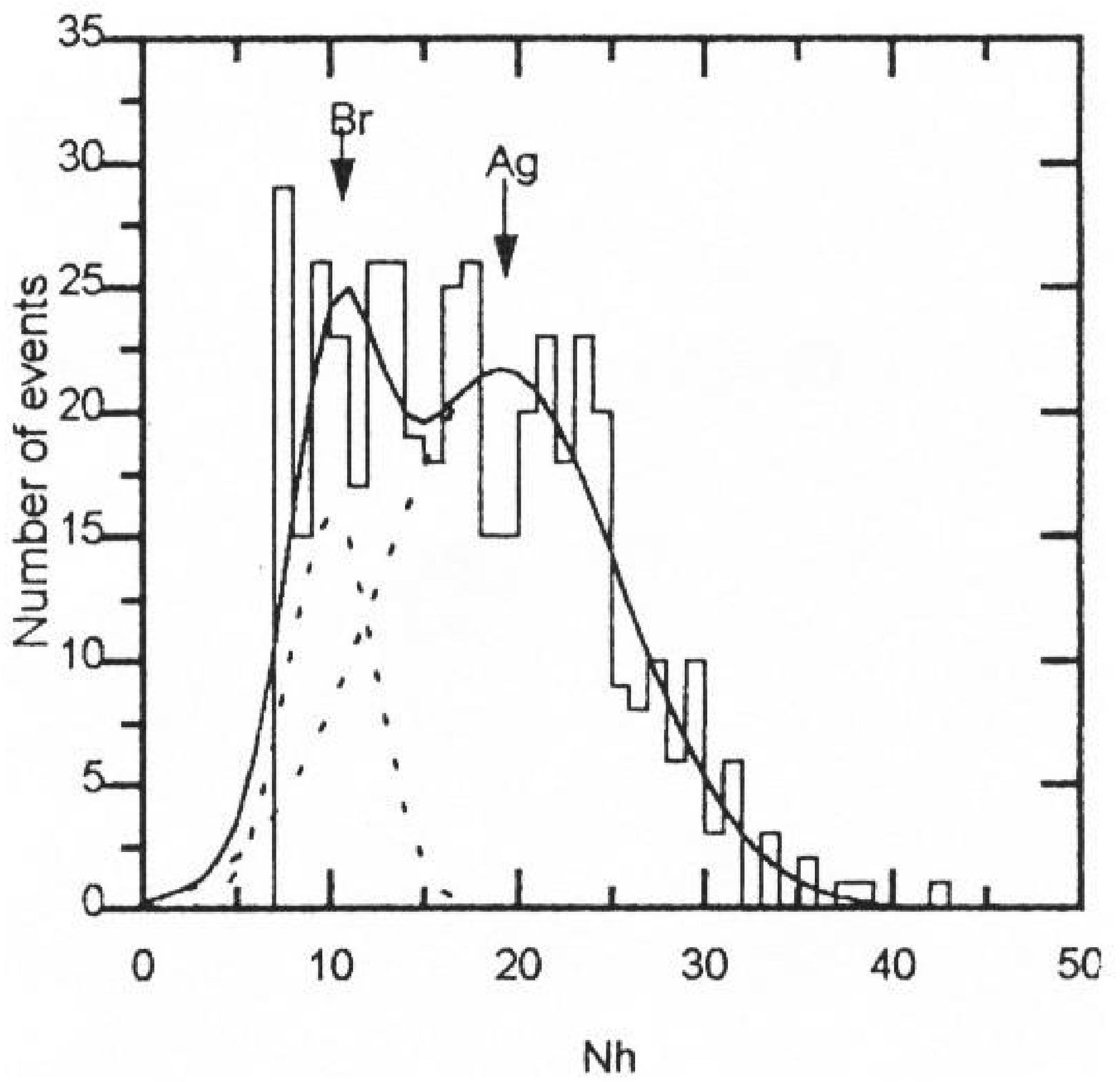}
\caption{
(a) Normalized multiplicity distribution of $N_{h}$ at $\sim$ 1 GeV per nucleon. Dotted line separate the events 
from the admixture of CNO target and peripheral collisions with Ag(Br) targets.
(b) $N_{h}$ distribution of events having $N_{h} \ge 8$. Solid line is a double Gaussian fit to separate Ag and 
Br target events. 
}
\label{ksnpssite}
\end{figure}

\section{Method of Target Identification}

The exact target identification in an emulsion experiment is not possible as the medium is composed 
of various elements as mentioned in {\bf table 2}. However, we can divide the major constituent elements into 
three broad target groups such as H (light), CNO (medium) and AgBr (heavy) with high accuracy. There 
are a lot of other ways of statistical separation {\bf [17-20]}, which roughly give the probability of 
interactions with different targets.
It is well known that the number of heavy particles, $N_{h}$ is a good tool for target identification. 
Since we are interested in the separation of targets on event by event basis, we have employed 
short-range track distribution to identify targets for events with low $N_{h}$ value. In view of the 
distribution of heavily ionizing charge particles for all set of events as shown in {\bf fig. 1(a)}, we have 
attempted the separation of targets using the following criteria:

{\bf H target events :} $N_{h}$ = 0; $N_{h}$ = 1 but not falling in any of the below categories.

{\bf CNO target events :} 2 $\le N_{h} \le $ 8 and no track with range $\ge$ 10 $\mu$m.

{\bf Ag(Br) target events :} $N_{h}$ $>$ 8; $N_{h}\le$ 8 and at least one track with range $\ge$ 10 $\mu$m 
and no track with 10 $\le$ range  $\le $ 50 $\mu$m.

As a result, we have obtained the percentage for the occurence of the three different target group 
events as H : 12$\%$, CNO: 48$\%$ and Ag(Br): 40$\%$. The relevant data is summarised 
in {\bf table 3}. Table 3 gives the results using the above criteria along with the results of other similar 
efforts {\bf [21-28]}. It may be seen from the table that the probability of events due to Ag(Br) nuclei increases slowly
with increasing projectile mass at similar energy. It also shows that the method of target separation is almost correct.

\begin{table}
\caption{Percentage of interactions with different target groups.}
\begin{tabular}{cccccc}
\hline
{\bf Interactions}&{\bf Energy} (A GeV)&{\bf H}&{\bf CNO}&{\bf Ag(Br)}& {\bf Ref.}\\

\hline
$^{1}P$&2.5&18.00&49.50&32.50&{\bf [21]}\\
$^{4}He$&3.7&21.03&40.42&38.55&{\bf [22]}\\
$^{12}C$&3.7&21.29&30.87&47.84&{\bf [23]}\\
$^{22}Ne$&3.7&12.94&32.59&54.47&{\bf [24]}\\
$^{28}Si$&3.7&15.29&33.79&50.92&{\bf [25]}\\
$^{40}Ar$&1.8&17.80&34.60&47.50&{\bf [26]}\\
$^{56}Fe$&3.7&23.13&22.64&54.23&{\bf [27]}\\
$^{84}Kr$&1.0&12.10&47.60&40.40&{\bf [Present work]}\\
$^{197}Au$&8.7.&19.00&36.00&45.00&{\bf [28]}\\
\hline
\end{tabular}
\end{table}

We are, first-ever, separating the Ag and Br nuclei in emulsion detector. For this, we examine only 
Ag(Br) type interaction. Their frequency distribution is shown in {\bf fig. 1(b)} and has been fitted by the 
double Gaussian functions. We have found that the contributions of Ag nuclei and of Br nuclei are 
44$\%$ and 56$\%$ ,respectively. The same numbers for Ag and Br nuclei are reported by 
B. Jakobsson et al {\bf [29]}, who employed the distribution of the sum of charges in events where all targets 
have been measured directly from an exposure at GANIL.
From now, we will proceed for development of new method for impact parameter estimation on event 
by event basis only for separate target group to avoid the mathematical complications and making 
the method simpler.

\begin{figure}
\center
\includegraphics[height=6.0cm,angle=0]{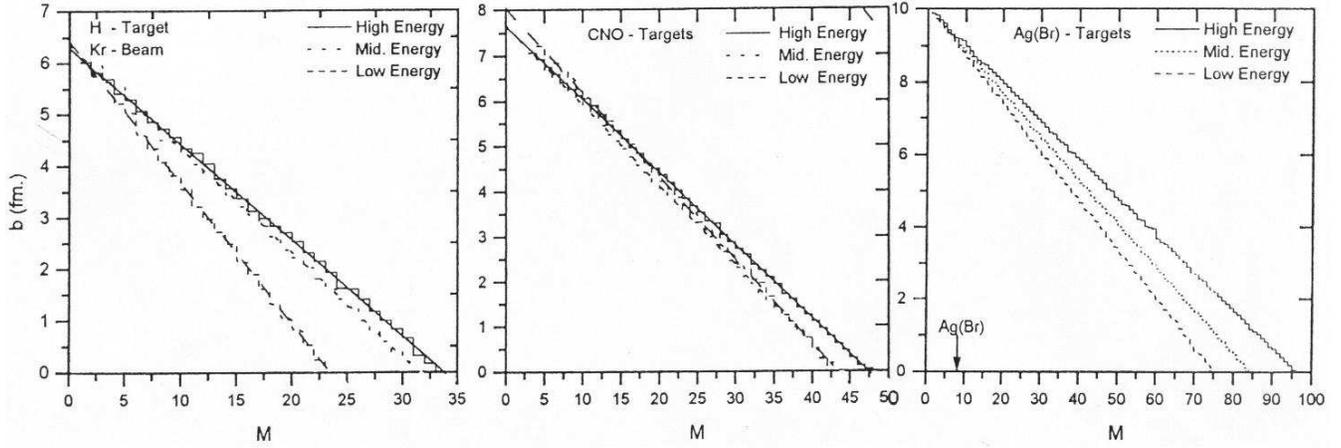}
\caption{
Reference distribution of b and total charge particle multiplicity for different target groups and for 
different energy intervals. Solid line is just to guide the eye.
}
\label{summaryplots}
\end{figure}

\section{Method of Impact parameter(b) Estimation}

Shower particle multiplicity have strong correlation with impact parameter in the fix target mass experiments. 
Due to the invariant target mass, shower particle multiplicity will not show strong correlation with impact 
parameter. Because H target, with fix mass beam, having nearly zero impact parameter may show the similar shower 
multiplicity as CNO or Ag(Br) target, with fix beam, having larger impact parameters. That's the reason we are 
interested in using total charged particle multiplicity ($N_{ch}$) to estimate the impact parameter and will 
show strong correlation between them. In some collider experiments {\bf [30]} they found that the total 
particle multiplicity ($N_{ch}$) is strongly correlated with the impact parameter b. More precisely, its mean 
value decreases monotonically as a function of b. 
  
In this method the basic assumptions are as following:
We treated interactions of each target group with projectile separately to make the method simpler. The total 
cross section is purely geometrical and there is a strong correlation between b and $N_{ch}$. The targets are 
randomly distributed in the nuclear emulsion detector and the probability of interactions is totally random and 
the collisions geometry is also random.

Therefore, we used random number generator {\bf [31]} to generate random numbers in between 0 and 1 for $N_{ch}$ 
and b. For making the random number values upto the order of total multiplicity ($N_{ch}$), we 
choose the maximum and minimum limit according to our real experimental values of maximum and minimum multiplicities 
of total charged particles in each target group and each target group multiplied by a suitable factor of numbers. 
For making the random number values upto the order of impact parameter according to the $A_{P}$ and $A_{T}$, we used 
real radius values of projectile and target nuclei and took the maximum and minimum as a combination and difference 
of radii of both nuclei, for b = $R_{P}$ + $R_{T}$ = $b_{max}$ and $b = | R_{P} \sim R_{T}| = b_{min} (\approx 0)$, 
respectively and made the order of magnitude according to the order of real experimental values. We assume that minimum 
multiplicity value belong to the $b_{max}$ and maximum multiplicity belongs to the $b_{min}$ and therefore we set 
two extreme points on the b versus multiplicity distribution and distributed rest of the values according to the 
monotonic relation between b and $N_{ch}$ such as b = $N_{ch(max)}$ - $N_{ch}$ for different target groups and for 
different energy intervals. In {\bf fig. 2}, we are showing relation between generated impact parameter and generated 
multiplicity of charged particles. We know the real event total charge particle multiplicity $N_{ch}$ and match this 
real multiplicity with b with the help of {\bf fig. 2} and can easily estimate the impact parameter corresponding 
to the real event.

\section{Experimental Results} 

To check the authenticity of the developed method, we have checked few charecteristic parameters with respect to impact 
parameter. Most of them are also checked in different energy intervals as well as different target groups. In this 
sequence, first we checked total charge particls multiplicity with respect to impact parameter for different target 
groups in different energy intervals as shown in {\bf fig. 3 (a), (b) and (c)}, respectively. The error bars shown 
on the data points are purely statistical and different type of lines are the best fit of data in different energy 
intervals. From fig. 3, we may infer that in each target group, the relation between multiplicity and impact parameter 
is monotonic and the slope is variable with change in beam energy as well as change in the target mass number. The nature 
of the distribution is similar to the participant - spectator Model prediction i.e., at maximum overlap ($b_{min}$) 
region of target and projectile, multiplicity must be maximum and vice - versa.    
\begin{figure}
\center
{\bf a)}
\includegraphics[height=4cm,angle=0]{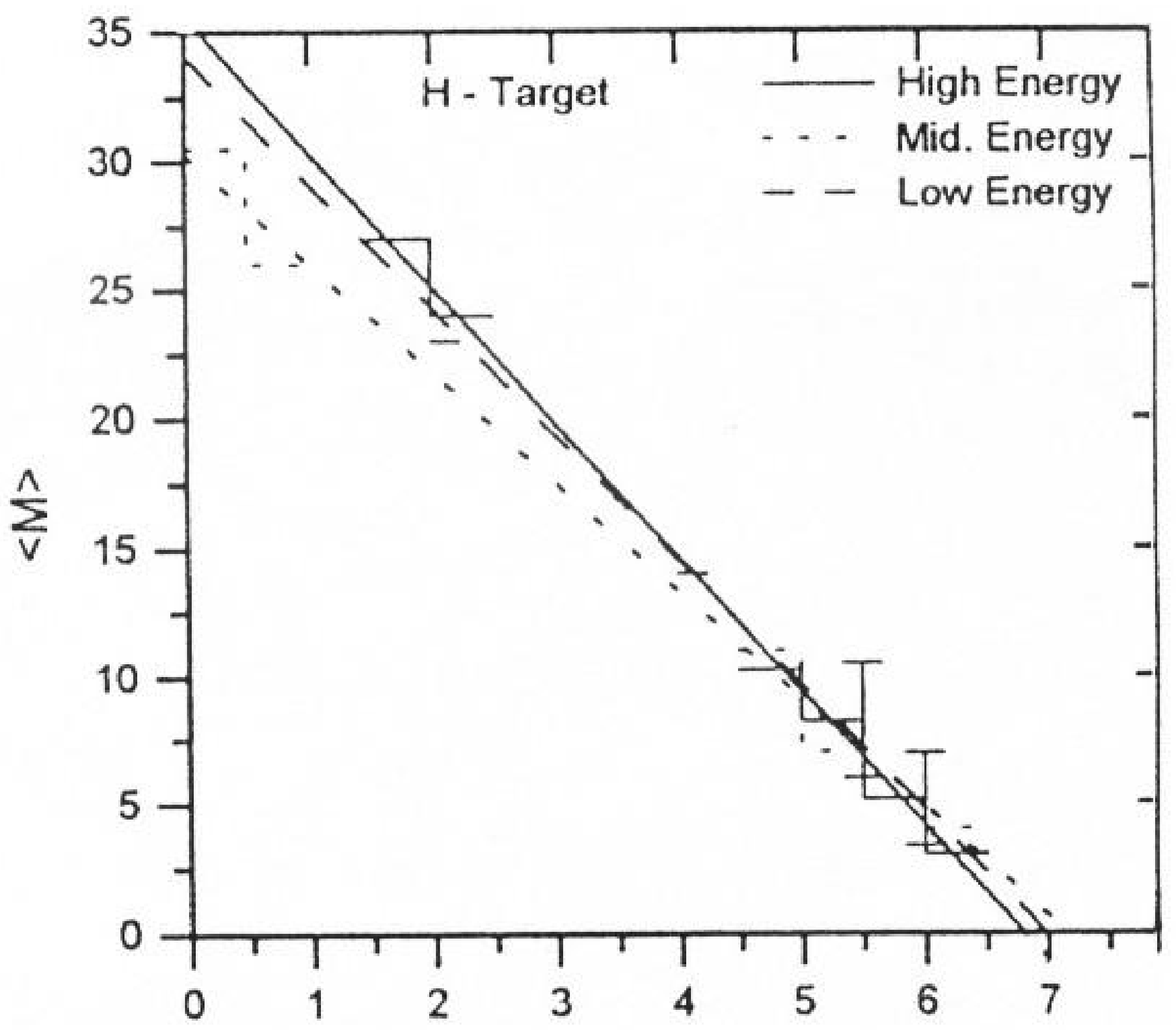}
{\bf b)}
\includegraphics[height=4cm,angle=0]{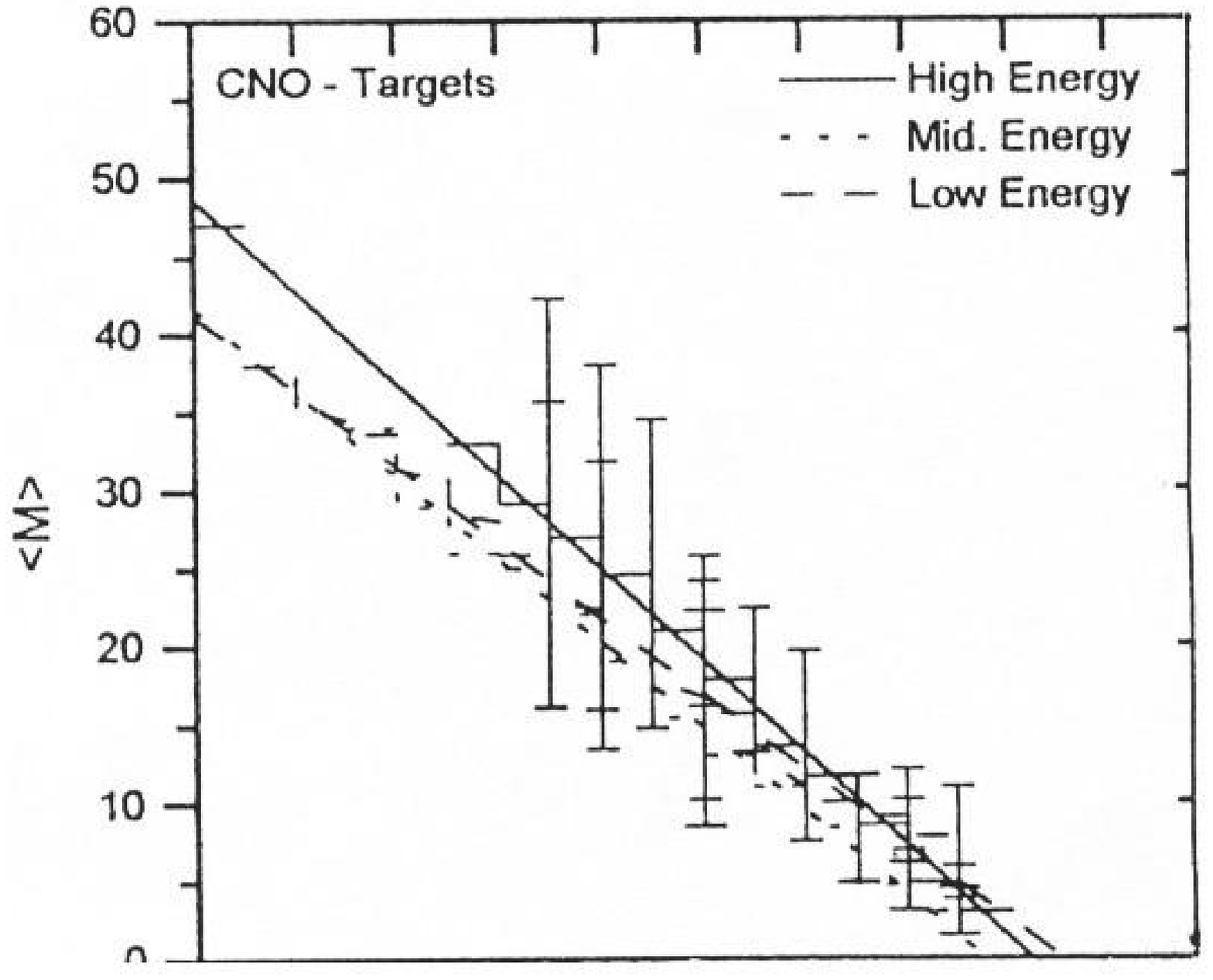}
{\bf c)}
\includegraphics[height=4cm,angle=0]{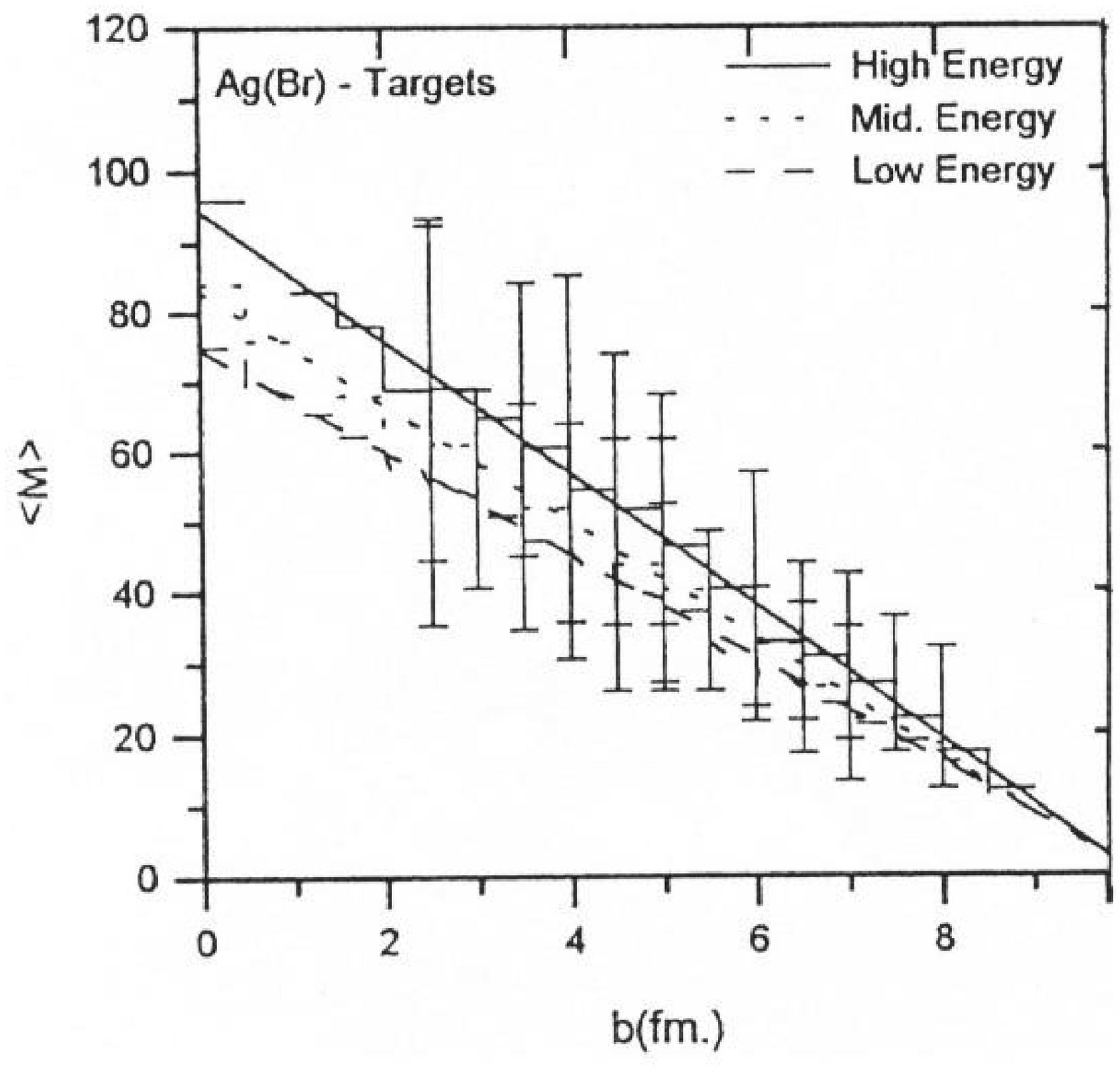}
\caption{
The average total charge multiplicity is plotted as a function of estimated impact parameter in different energy 
intervals for different target groups {\bf (a)} H, {\bf (b)} CNO, and {\bf (c)} Ag(Br).Solid, short dash and long 
dash lines are the best fit of high energy, mid. energy and low energy data set, respectively. 
}
\label{ksnpssite}
\end{figure}

We checked the emission of alpha particles in interactions with H - target with respect to the impact parameter 
as shown in {\bf fig. 4(a)}. The distribution shows a polynomia nature. The maximum number of emitted alpha particle 
in an interaction with H - target is 5 to 6 at maximum overlap. Here the size of H - target is much smaller than 
the size of the $^{84}Kr$- projectile. That's why it is very difficult to select head-on collisions and this can 
easily mix into the central as well as quasi-central events.
Emission of the average number of shower particle versus b is plotted in {\bf fig. 4(b)}. The condition of this 
distribution is similar to the {\bf fig. 4(a)}. It shows a strong monotonic correlation between $<N_{s}>$ and b. 
The solid line is the best fit of data points and the error bars represent the statistical error.

\begin{figure}
\center
{\bf a)}
\includegraphics[height=7cm,angle=0]{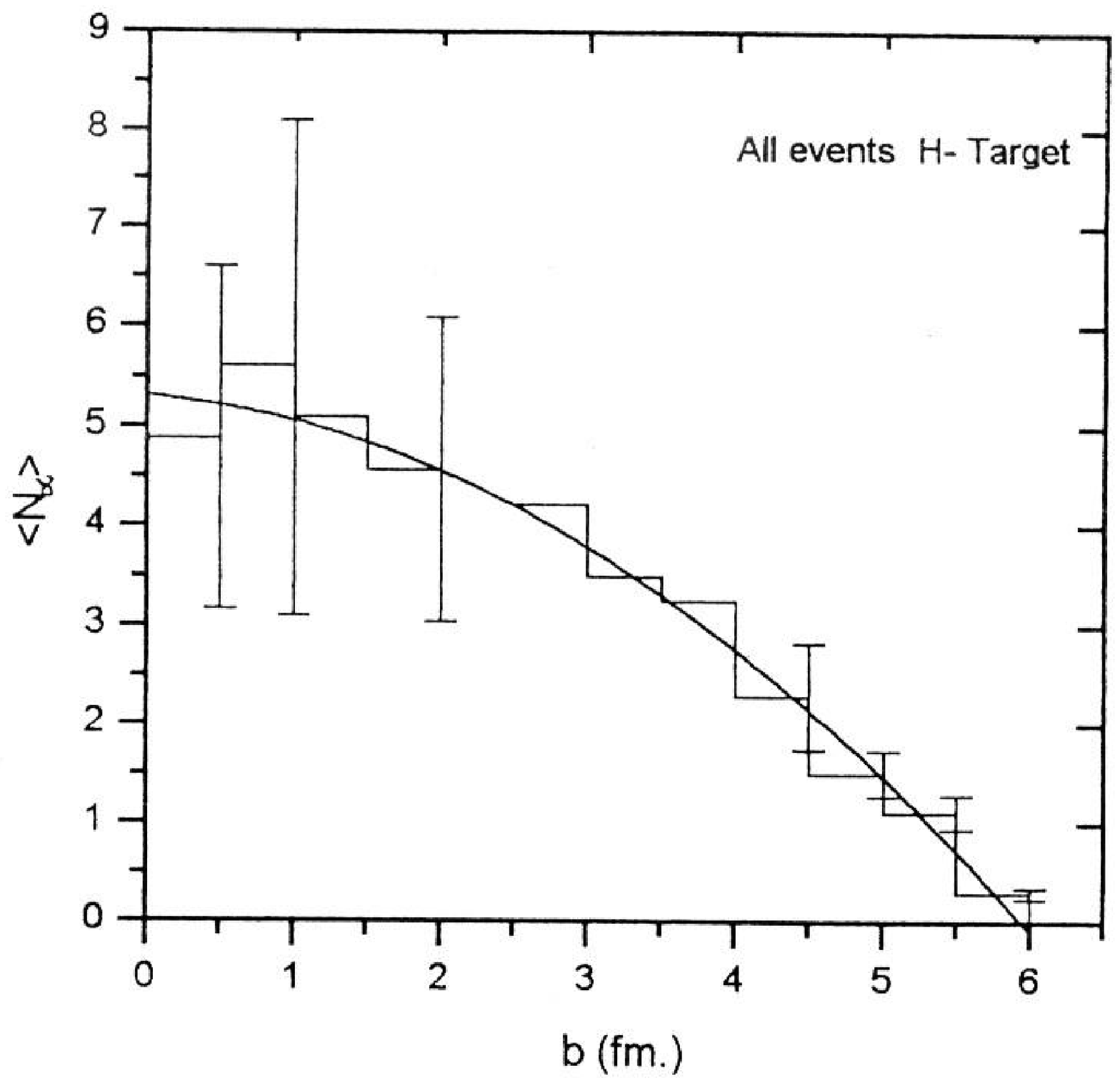}
{\bf b)}
\includegraphics[height=7cm,angle=0]{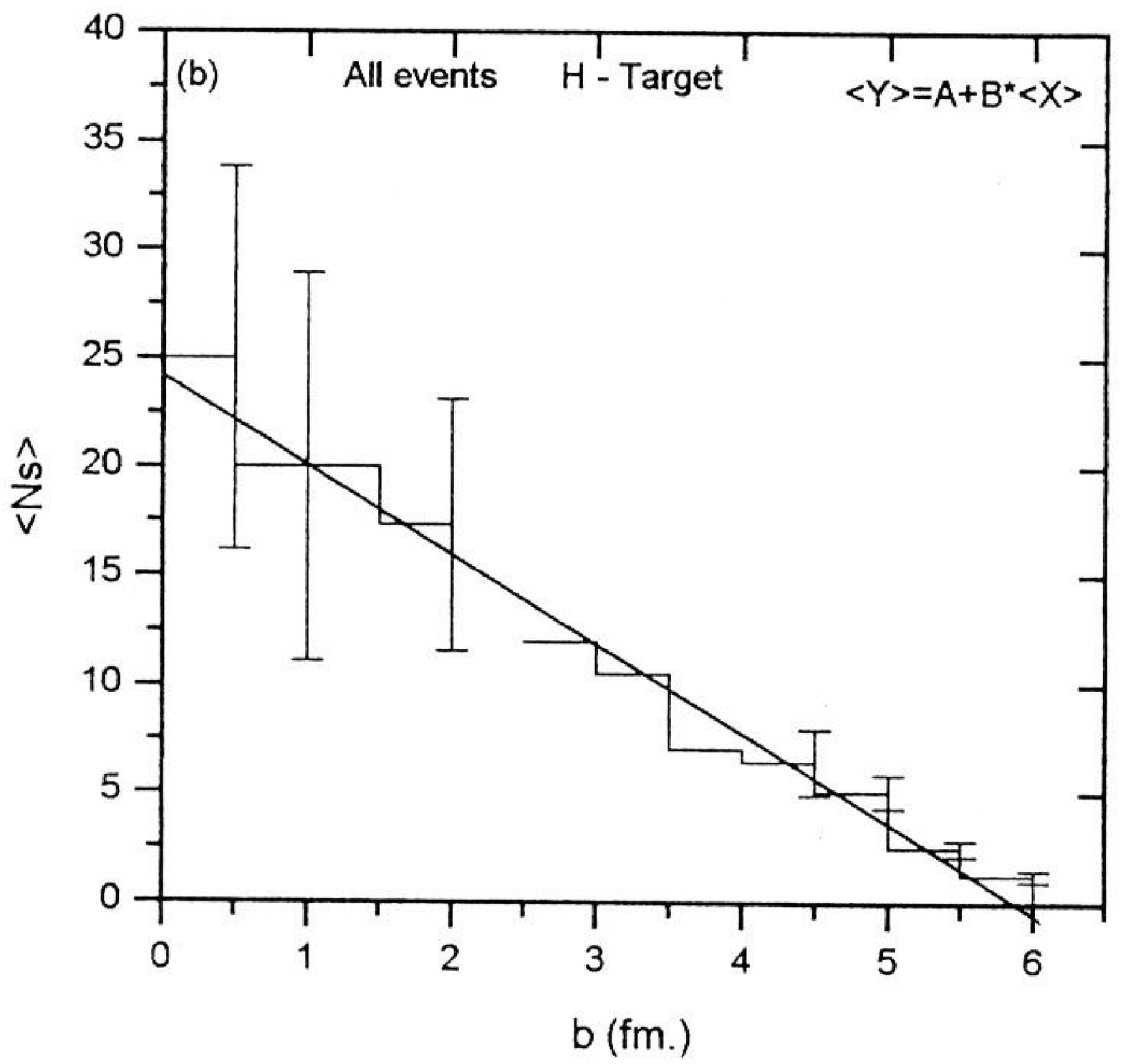}
\caption{
(a) Average number of Helium nuclei emitted in interaction with H - target are plotted with respect to impact parameter 
for all energy. Solid line is the polynomial function fit is just to guide the eye.  
(b) Average number of shower particles emitted in interaction with H - target are plotted as a function of impact parameter 
for all energy range (0.06 - 0.95 A GeV). Solid line is the best line fit of data points.
}
\label{ksnpssite}
\end{figure}

The {\bf Q (= $\Sigma f_{Z=1} + 2 \times \Sigma f_{Z=2} + \Sigma Z\times f_{Z\ge3}$)} is the total projectile 
fragments charge flow in the forward emission cone. The $<Q>$ value versus b for all energy events in different 
target groups is shown in {\bf fig. 5}. For making a comparison and guiding the eye, we have fitted the distribution 
with a linear function. The light and heavy target groups show nearly similar slopes while medium target group 
has slightly different slope. We can also see the $<Q>$ value change with target mass, specially with light 
and heavy targets. At $b_{min}$ and $b_{max}$, the difference in $<Q>$ is around double. At $b_{min}$, less 
than half of the projectile mass are converted into some other things (like energy and neutral particle) but 
at $b_{max}$ nothing is going to disappar while few neutrons are converted into protons and make the total 
charge {\bf Q} value larger than 36.

\begin{figure}
\center
{\bf }
\includegraphics[height=8cm,angle=0]{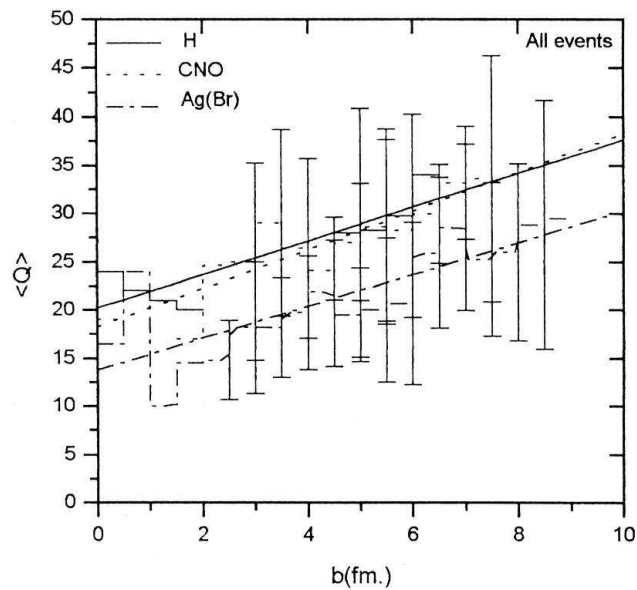}
\caption{
The average Q values versus impact parameter for high energy interval with different target groups. Different 
lines are best fit of the data points and are guiding to the eye.
}
\label{ksnpssite}
\end{figure}

In {\bf fig. 6}, we have plotted $<b>$ versus Q for all high energy events for one combined target 
(H + CNO + Ag and Br). It can be seen that Q is minimum at minimum value of $<b>$ and approaches the beam 
charge (Z = 36) at maximum value of $<b>$. The presence of some events having Q values more than the value 
of beam charge (36) when $<b>$ has a value less than the maximum $<b>$ may be attributed to the conversion 
of neutrons into protons while nuclei collide with each other. All above described behaviour of paramerets 
are consistent to the several theoretical models {\bf [32]}specially participant spectator model {\bf [33]}.

\begin{figure}
\center
{\bf}
\includegraphics[height=8cm,angle=0]{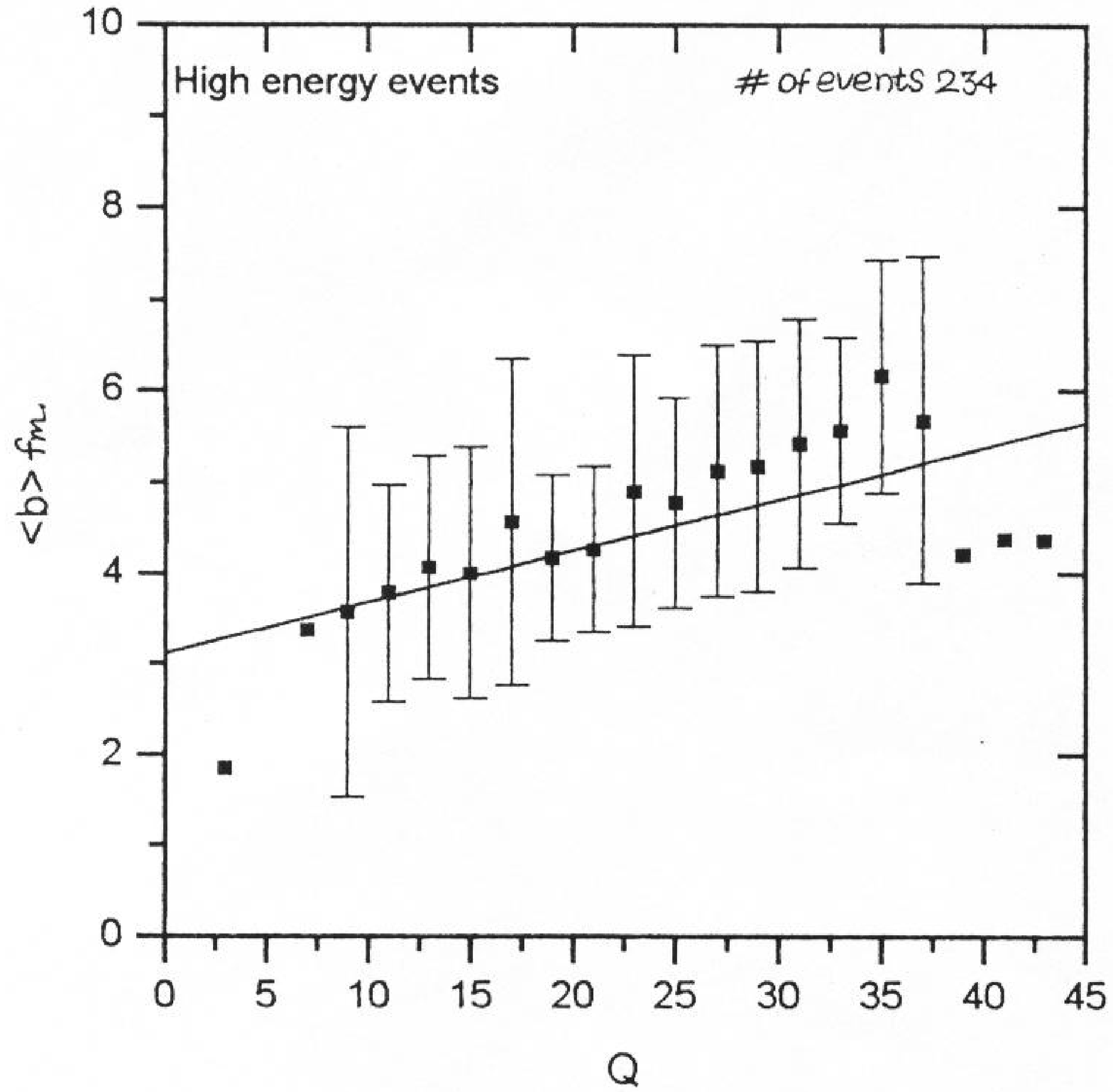}
\caption{
The average impact parameter versus Q for high energy interval. Solid line is the best fit of data points and 
is just to guide the eye.
}
\label{ksnpssite}
\end{figure}

\section{Conclusions}

A new method to estimate an Impact parameter on the basis of event-by-event in Photographic Nuclear Emulsion 
detector, was developed and studied in detail of some characteristic parameter. Although this work originates 
from the goals of handling data from photographic nuclear emulsion detector,  
and can be applicable to the similar kind of detectors or other multi-target detector systems where all 
type of targets are mixed together and only very limited information is in hand. These studies will boost 
the data analysis for emulsion detector in the light of impact parameter. This is a new method that's why 
we have no any other experimental data for comparison but the experimental values obtained in this experiments 
strongly support the theorectical concepts.

\section{Acknowledgments}

The authors would like to dedicate this work to Prof. S.K. Tuli, Fmr. Head of Physics Department, Leading Member 
of Experimetnal High Energy Physics Group, Banaras Hindu University, Varanasi, India. One of the author (VS) is 
thankful to Dr. B.K. Singh, Institute of Physics, Academia Sinica, Taipei, Taiwan for helpful comments and discussions 
regarding the subject matter. We are also grateful to the authorities of GSI, Darmstadt, Germany for their help 
in exposure of the emulsion stacks.


\begin{thebibliography}{99}
\bibitem{texono}
L.P. Csernai and J.I. Kapusta, 
Rep., 131, 223 (1986);
S.Nagamiya and M. Gyulassy, 
Adv. Nucl. Phys., 13, 201 (1982);
R. Stock, 
ibid., 135, 259 (1986);
H. Stoecker and W. Greiner, 
ibid., 137, 277 (1986).
\bibitem{start}
D. L'Hote,
Nucl. Phys. {\bf A488}, 457c (1988).
\bibitem{sciencemag}
H.H. Gutbrod, A.M. Poskanzer and H.G. Ritter,
Rep. Prog. Phys. 52, 1267 (1989).
\bibitem{pdg}
C. Cavata et al.,
Phys. Rev. C42, 1760 (1990).
\bibitem{prospects}
W.H. Barkas, Nuclear Research Emulsion, Vol. 1, Academic Press, New York, London (1963). 
\bibitem{program}
S. Nagamiya, Proc. of Int. Conference on Nuclear Physics, Bombay (1984);
U. Nantes, IRES Strasbourg and LEPSI Strabourg ``Proposal for a Silicon Strip Detector for STAR'' (1997);
VECC, Calcutta; Inst. Physics, Bhubaneshwar; Rajesthan Univ., -Jaipur; Punjab Univ.,-Chandigarh, and 
Jammu Univ.,-Jammu ``Proposal for a Photon Multiplicity Detector in STAR'' (1996).
\bibitem{eledaq}
P. Powell, F. Fowler and D.H. Perkins, A study of Elementary Particles by Photographic Mehtod, 
Pergamon Press, London (1959).
\bibitem{sigmanue}
V. Singh,
Ph.D. Thesis, Physics Department, Banaras Hindu University, Varanasi, India (1998).
\bibitem{}
H.H. Heckman et al., Phys. Rev. C17, 1735 (1978).
\bibitem{}
B. Judek, in Proceedings of the Fourteenth International Conference on Cosmic Rays, Munich, 
1975, edited by Klaus Pinkau (Max-Planck-Institute, Munchen, Vol.7, p. 2342 (1975).
\bibitem{}
L.K. Mangotra et al., IL Nuovo Cimento 87A, 279 (1985).
\bibitem{}
A. Gill et al., Int. J. of Mod. Phys. A5, 755 (1990).
\bibitem{}
C.J. Waddington and P.S. Freier, Phys. Rev. C31, 888 (1985).
\bibitem{}
E.M. Friedlander et al., Phys. Rev. C27, 2436 (1983).
\bibitem{}
A. Abdelsalam et al., J. Phys. G: Nucl. Part. Phys. 28, 1375 (2002). 
\bibitem{dyrange}
S.A. Krasnov et al., Czech. J. of Phys. 46, 531 (1996).
\bibitem{sensit}
J. Babecki, Act. Phys. Pol. B6, 443 (1975).
\bibitem{munupaper}
E.V. Anzon et al., Sov. J. Nucl. Phys. 22, 380 (1976).
\bibitem{rdk}
A. Abdelsalam, JINR Report, El-81-623 (1981).
\bibitem{proto}
I. Otterlund, LUIP-CR-76-05 (1976).
\bibitem{csibkg}
M. Bogdanski et al., Helv. Phys. Acta 42, 485 (1969). 
\bibitem{}
DGKLMTW collaboration, JINR Dubna Commun., P1-8313 (1974).
\bibitem{lens}
T. Ahmad, Ph.D. Thesis, Aligarh Muslim University, Aligarh, India (1991).
\bibitem{gso}
Z. Abou-Moussa, Can. J. Phys. 80, 109 (2002).
\bibitem{naicdm}
B.K. Singh et al., Nucl. Phys. A570, 819 (1994).
\bibitem{nbeam}
R.R. Joseph, Ph.D. Thesis, Banaras Hindu University, Varanasi, India (1986).
\bibitem{lepsd}
R. Bhanja et al., Nucl. Phys. A411, 507 (1983).
\bibitem{kims}
A. Dabrowska et al., Z. Phys. C59, 399 (1993).
\bibitem{amsradio}
B. Jakobson and R. Kullberg, Phys. Scr. 13, 327 (1976).
\bibitem{ciaeams}
K. Adcox et al.(Phenix Collaboration), Phys. Rev. Lett. 86, 3500 (2001); 
K. Adcox et al.(Phenix Collaboration), Phys. Rev. Lett. 052301 (2001); 
K. Adcox et al.(Phenix Collaboration), Phys. Rev. Lett 88, 192302 (2002);
K. Adcox et al.(Phenix Collaboration), Phys. Lett. B561, 82 (2003);
J. Adams et al.(Star Collaboration), Phys. Rev. Lett.92, 182301 (2004).  
\bibitem{Spring8}
Portable random number generator proposed by L'Ecuyer in Comm. ACM 31:743 (1988). 
\bibitem{Spring8}
I. Otterlund, Proc. 4th High Energy Heavy Ion Summer Study, Berkeley LBL-7766, 289 (1978);
M.I. Adamovich et al., J. Phys. G: Nucl. Part. Phys. 22, 1469 (1996);
J. Hufner, GSI Priprint 80-1 91880);
J. Hufner et al., Nucl. Phys. A290, 460 (1977);
A. Bialas et al., Phys. Rev. D25, 2328 (1982);
K. Kinoshita et al., Z. Phys. C8, 205 (1981).
\bibitem{Spring8}
J. Knoll et al., Nucl Phys. A308, 500 (1978);
M. Guylassy et al., Phys. Rev. Lett. 40, 298 (1978);
G.D. Westfall et al., Phys. Rev. Lett. 37, 56 (1976).
\end{thebibliography}
\end{document}